\documentclass[traditabstract,longauth]{aa}  

\usepackage{natbib}
\bibpunct{(}{)}{;}{a}{}{,}
\usepackage{graphicx}
\usepackage{txfonts}

\newcommand{\ngc}{NGC$\,1097$}
\newcommand{\spitzer}{\textit{Spitzer}}

\newcommand \msun{\hbox{$\hbox{M}_{\odot}$}}
\newcommand{\micron}{$\mu$m}

\begin{document}

\title{Mapping far-IR emission from the central
kiloparsec of \ngc\thanks{Herschel is an ESA space observatory with
science instruments provided by European-led Principal Investigator
consortia and with important participation from NASA.}}

\author{K.\ Sandstrom\inst{1} 
\and O.\ Krause\inst{1}     
\and H.\ Linz\inst{1} 
\and E.\ Schinnerer\inst{1}   
\and G.\ Dumas\inst{1}      
\and S.\ Meidt\inst{1} 
\and H.-W.\ Rix\inst{1}       
\and M. Sauvage\inst{2}          
\and F.\ Walter\inst{1} 
\and R. C.\ Kennicutt\inst{3} 
\and D.\ Calzetti\inst{4}
\and P.\ Appleton\inst{5}     
\and L.\ Armus\inst{6}      
\and P.\ Beir\~{a}o\inst{6}
\and A.\ Bolatto\inst{7}      
\and B.\ Brandl\inst{8}     
\and A.\ Crocker\inst{4}
\and K.\ Croxall\inst{9}      
\and D.\ Dale\inst{10}      
\and B. T.\ Draine\inst{11}
\and C.\ Engelbracht\inst{12} 
\and A.\ Gil de Paz\inst{13}
\and K.\ Gordon\inst{14}      
\and B.\ Groves\inst{8}     
\and C.-N.\ Hao\inst{15}
\and G.\ Helou\inst{5}        
\and J.\ Hinz\inst{12}      
\and L.\ Hunt\inst{16}
\and B. D.\ Johnson\inst{3}      
\and J.\ Koda\inst{17}      
\and A.\ Leroy\inst{18}
%\and K.\ Misselt\inst{12}     
\and E. J.\ Murphy\inst{6}     
\and N.\ Rahman\inst{7}
\and H.\ Roussel\inst{19}     
\and R.\ Skibba\inst{12}    
\and J.-D.\ Smith\inst{9}
\and S.\ Srinivasan\inst{19}  
\and L.\ Vigroux\inst{19}   
\and B. E.\ Warren\inst{20}
\and C. D.\ Wilson\inst{21}   
\and M.\ Wolfire\inst{7}    
\and S.\ Zibetti\inst{1}}

\institute{Max-Planck-Institut f\"{u}r Astronomie, K\"{o}nigstuhl 17,
69117 Heidelberg, Germany \email{sandstrom@mpia.de}
\and CEA/DSM/DAPNIA/Service d'Astrophysique, UMR AIM, CE Saclay, 91191
Gif sur Yvette Cedex
\and Institute of Astronomy, University of Cambridge, Madingley Road,
Cambridge CB3 0HA, UK
\and Department of Astronomy, University of Massachusetts, Amherst, MA
01003, USA
\and NASA Herschel Science Center, IPAC, California Institute of
Technology, Pasadena, CA 91125, USA
\and Spitzer Science Center, California Institute of Technology, MC
314-6, Pasadena, CA 91125, USA
\and Department of Astronomy, University of Maryland, College Park, MD
20742, USA
\and Leiden Observatory, Leiden University, P.O. Box 9513, 2300 RA
Leiden, The Netherlands
\and Department of Physics and Astronomy, University of
Toledo, Toledo, OH 43606, USA
\and Department of Physics \& Astronomy, University of Wyoming,
Laramie, WY 82071, USA
\and Department of Astrophysical Sciences, Princeton University,
Princeton, NJ 08544, USA
\and Steward Observatory, University of Arizona, Tucson, AZ 85721, USA
\and Departamento de Astrofisica, Facultad de Ciencias Fisicas,
Universidad Complutense Madrid, Ciudad Universitaria, Madrid, E-28040,
Spain
\and Space Telescope Science Institute, 3700 San Martin Drive,
Baltimore, MD 21218, USA
\and Tianjin Astrophysics Center, Tianjin Normal University, Tianjin
300387, China
\and INAF - Osservatorio Astrofisico di Arcetri, Largo E. Fermi 5,
50125 Firenze, Italy
\and Department of Physics and Astronomy, SUNY Stony Brook, Stony
Brook, NY 11794-3800, USA
\and Hubble Fellow, National Radio Astronomy Observatory, 520 Edgemont
Road, Charlottesville, VA 22903, USA
\and Institut d'Astrophysique de Paris, UMR7095 CNRS, Universit\'e
Pierre \& Marie Curie, 98 bis Boulevard Arago, 75014 Paris, France
\and ICRAR, M468, University of Western Australia, 35 Stirling Hwy,
Crawley, WA 6009 Australia
\and Department of Physics \& Astronomy, McMaster University,
Hamilton, Ontario L8S 4M1, Canada}

%%\date{}

\abstract{Using photometry of NGC 1097 from the {\em Herschel} PACS
(Photodetector Array Camera and Spectrometer) instrument, we study the
resolved properties of thermal dust continuum emission from a
circumnuclear starburst ring with a radius $\sim 900$ pc.  These
observations are the first to resolve the structure of a circumnuclear
ring at wavelengths that probe the peak (i.e.  $\lambda \sim 100$
\micron) of the dust spectral energy distribution.  The ring dominates
the far-infrared (far-IR) emission from the galaxy---the high angular
resolution of PACS allows us to isolate the ring's contribution and we
find it is responsible for 75, 60 and 55\% of the total flux of NGC
1097 at 70, 100 and 160 \micron, respectively.  We compare the far-IR
structure of the ring to what is seen at other wavelengths and
identify a sequence of far-IR bright knots that correspond to those
seen in radio and mid-IR images.  The mid- and far-IR band ratios in
the ring vary by less than $\pm 20$\% azimuthally, indicating modest
variation in the radiation field heating the dust on $\sim 600$ pc
scales.  We explore various explanations for the azimuthal uniformity
in the far-IR colors of the ring including a lack of well-defined age
gradients in the young stellar cluster population, a dominant
contribution to the far-IR emission from dust heated by older ($> 10$
Myr) stars and/or a quick smoothing of local enhancements in dust
temperature due to the short orbital period of the ring.  Finally, we
improve previous limits on the far-IR flux from the inner $\sim 600$
pc of NGC 1097 by an order of magnitude, providing a better estimate
of the total bolometric emission arising from the active galactic
nucleus and its associated central starburst.}

   
\maketitle

\section{Introduction}

The central regions of galaxies host some of the most intense
star-formation that we can observe in the local Universe in
circumnuclear starburst rings.  Starburst rings are believed to be the
consequence of the pile-up of inflowing gas and dust, driven by a
non-axisymmetric potential from a stellar bar, on orbits located near
the Inner Lindblad Resonance of the bar
\citep{combes85,athanassoula92}.  The high surface densities that
exist in the ring lead to high star-formation rates. Indeed starburst
rings are one of few regions in non-interacting galaxies where the
formation of ``super star clusters'' commonly occurs \citep{maoz96}.
The stars formed in the ring can be numerous enough to drive the
structural evolution of the galaxy \citep{norman96,kormendy04} and can
be the dominant power source for the galaxy's infrared (IR) emission.  

Star-formation in circumnuclear rings occurs under conditions not
normally found in the disks of galaxies: in addition to their high gas
surface densities, these regions have dynamical timescales that are
comparable to the lifetimes of massive stars.  Understanding
star formation in circumnuclear rings has been a long-standing problem
\citep{combes96}.  There are two main models: the ``popcorn'' model
\citep{elmegreen94}, where star-formation is driven by stochastic
gravitational fragmentation along the ring, and the ``pearls on a
string'' model, where gas flowing into the ring is compressed near the
contact points (i.e. locations where the dust lanes intersect the
ring) and then forms stars a short distance downstream
\citep[e.g.,][]{boker08}.  The ``pearls on a string'' model predicts a
gradient in the ages of young stellar clusters as one moves away from
the contact points.  This has been observed in a number of starburst
rings \citep[e.g., ][]{mazzuca08,boker08}.  Conversely, many
well-studied rings show no evidence for an age gradient
\citep{maoz01}.  It is not obvious, however, that a single mode of
star-formation must occur in all rings or even at all times in a given
ring \citep{vandeven09}.   

KINGFISH (Key Insights into Nearby Galaxies: A Far-Infrared Survey
with {\em Herschel}, PI R. Kennicutt) is an Open-Time Key Program to
study the interstellar medium (ISM) of nearby galaxies with
far-IR/sub-mm photometry and spectroscopy.  Among the unique aspects
of the KINGFISH science program is the ability to observe thermal dust
emission at unprecedented spatial resolution ($\sim$ 5.6, 6.8 and
11.3\arcsec at 70, 100 and 160 \micron) using PACS (Photodetector
Array Camera and Spectrometer) imaging.  High spatial resolution
is crucial for observing processes occurring in the central regions of
galaxies.  These regions represent our best opportunity to study in
detail the interplay between dynamics, star-formation and feedback
that regulate the fueling of nuclear activity, be it a starburst or an
active galactic nucleus (AGN).  

Below we present PACS imaging of the galaxy NGC 1097, one
of the first KINGFISH targets observed during the {\em Herschel}
Science Demonstration Program (SDP) (for PACS spectroscopy of NGC 1097
see Beir\~{a}o et al. 2010 and for {\em SPIRE} observations see
Engelbracht et al. 2010).  The source NGC 1097 is a barred spiral
galaxy located at a distance of 19.1 Mpc \citep[][1\arcsec$\approx 92$
pc]{willick97}.  In its central kpc it hosts an intensely star-forming
\citep[$\sim 5$ \msun\ yr$^{-1}$;][]{hummel87} ring with a radius of
$\sim 900$ pc.  The ring's rotation speed of $\sim 300$ km s$^{-1}$
\citep[corrected for inclination,][]{storchi-bergmann96}, corresponds
to a rotation period of $\sim 18$ Myr.  The galaxy's nucleus is
classified as a LINER from optical emission line diagnostics
\citep{phillips84}, but is shown to be a Seyfert 1 by its double-peaked
H$\alpha$ profile \citep{storchi-bergmann93}.  UV spectroscopy has
revealed a few Myr old burst of star-formation in the central 9
pc of the galaxy \citep{storchi-bergmann05}.  With the high spatial
resolution of {\em Herschel} PACS, we can resolve the starburst ring
and inner 600 pc of NGC 1097 for the first time at wavelengths near
the peak of the dust spectral energy distribution (SED).

\section{Observations and data reduction}

The galaxy NGC 1097 was observed with the PACS instrument (Poglitsch
et al. 2010) on the {\em Herschel} Space Observatory (Pilbratt et al.
2010) on 2009 December 20 during the SDP. We obtained 15\arcmin\ long
scan-maps in two orthogonal directions at the medium scan speed
(20\arcsec s$^{-1}$).  The scan position angles (45$^\circ$ relative
to the scan direction) provide homogeneous coverage over the mapped
region. The total on-source times per pixel were approximately 150,
150, and 300 seconds for 70, 100 and 160 \micron, respectively.  

The raw data were reduced with HIPE (Ott 2010), version 3.0, build
455. Besides the standard steps leading to level-1 calibrated data,
second-level deglitching and correction for offsets in the detector
sub-matrices were performed. The data were then highpass-filtered
using a median window of 5\arcmin\ to remove the effects of
bolometer sensitivity drifts and 1/f noise.  We masked out emission
structures (visible in a first iteration of the map-making) with a
5\arcmin-wide mask before computing this running median to
prevent subtraction of source emission.  Although the filtering may
remove some extended flux from the galaxy, because we are primarily
interested in the very bright central 1\arcmin\ of NGC 1097 this
effect will be negligible.  Finally, the data were projected onto
a coordinate grid with 1\arcsec\ pixels. 

\begin{figure*}
\begin{center}
\scalebox{0.47}{{\includegraphics{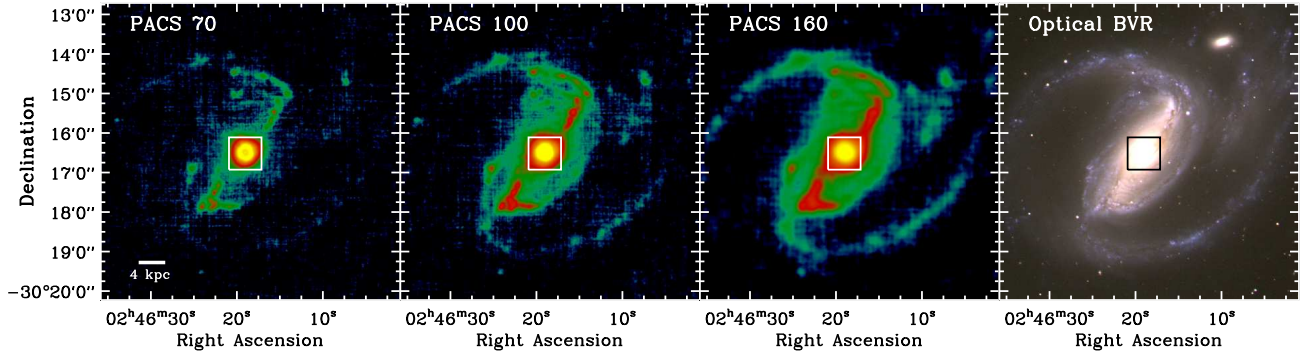}}}
\caption{PACS observations of NGC 1097 and an optical BVR image
from the SINGS ancillary observations \citep{kennicutt03} for
comparison.  The 70, 100 and 160 \micron\ bands are shown with a
logarithmic stretch that highlights the spiral arms.  The white square
in each panel shows the location of the region presented in
Fig~\ref{fig:ring}.\label{fig:rgb}}
\end{center}
\end{figure*}

After pipeline processing we applied flux correction factors from the
PACS team to adjust the calibration.  The current calibration has
uncertainties of $\sim 10$, 10, and 20\% for the 70, 100 and 160
\micron\ bands, respectively (Poglitsch et al. 2010).  Because we aim to
compare our PACS observations with ancillary data at other
wavelengths, we adjusted the relative astrometry of the PACS
observations to match that of the \spitzer\ 24 \micron\ from SINGS
\citep[Spitzer Infrared Nearby Galaxies Survey:][]{kennicutt03}.
This was done by measuring the positions of background point-sources
in the MIPS 24 \micron\ (Multi-Band Imaging Photometer) and PACS
100 \micron\ images, adjusting the PACS 100 \micron\ astrometry,
assuming the relative astrometry for the PACS bands is well-calibrated
and transferring the solution to the other bands.  The offset between
the PACS and MIPS astrometry was $\sim 2$\arcsec.  The one-sigma
surface brightness sensitivities per pixel of the final maps are 5.9,
6.2 and 3.3 MJy sr$^{-1}$.  In Fig~\ref{fig:rgb} we show the three
PACS images with a logarithmic stretch to highlight the spiral arms.
Note that below we extract photometry from the images
at their native resolution using apertures larger than the beam size
of the lowest resolution map.

\section{{\textit{Herschel} PACS observations of the circumnuclear
ring in NGC 1097}}

\begin{figure*}
\begin{center}
\scalebox{0.6}{{\includegraphics{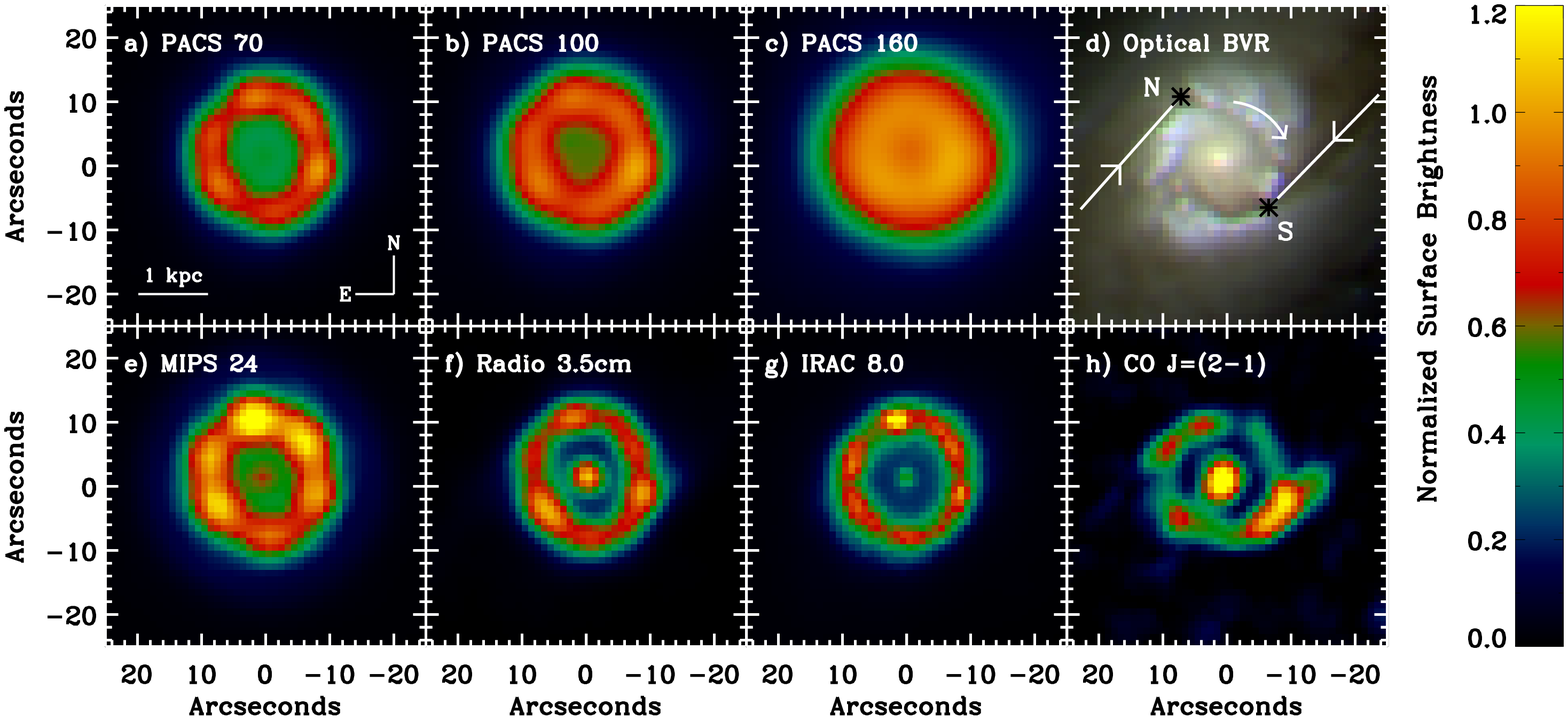}}}
\caption{Circumnuclear ring of NGC 1097 at a variety of
wavelengths. Each image is shown at its native resolution: $\sim 6$, 7
and 12\arcsec\ for PACS 70, 100 and 160; $\sim 6$\arcsec\ for MIPS 24;
$\sim 4$\arcsec\ for the 3.5 cm radio continuum \citep{beck05} and CO
J=(2-1) \citep{hsieh08}; $\sim 2$\arcsec\ for IRAC 8.0 \micron; and
$\sim 1$\arcsec\ for the optical B, V, and R imaging
\citep{kennicutt03}. All images, except the three-color BVR image, are
normalized to the surface brightness of the ring at the location
($-$8\arcsec,$-$1\arcsec), i.e. the brightest knot in the PACS 70
image.  The approximate locations of the dust lanes that feed the
circumnuclear ring are overlayed on the BVR three-color image and the
locations of the contact points are identified with black asterisks
and labeled with ``N'' and ``S''.  Material in the ring rotates
clockwise as shown with the arrow in panel h. Note that at this
stretch the central source is not visible at 70
\micron.\label{fig:ring}}
\end{center}
\end{figure*}

The most prominent far-IR structure in NGC 1097 is its circumnuclear
starburst ring, shown in Fig~\ref{fig:ring} at a variety of
wavelengths.  The PACS angular resolution allows us to clearly
separate the contribution of the ring and nucleus from the galaxy's
emission for the first time at wavelengths that probe the peak of the
dust SED.  Summing the emission within a radius of 20\arcsec (1.8 kpc)
of the center and comparing it with the total flux from the galaxy, we
find that the ring and nucleus emit 75, 60 and 55\% of the total flux
of NGC 1097 at 70, 100 and 160 \micron, respectively (there
is some galactic emission within 1.8 kpc that is not associated with
the ring or nucleus, but this component is negligible).  These
measurements imply that the SED of the more extended galactic emission
peaks at longer wavelengths than the SED of the ring.  Indeed, by
fitting a modified blackbody to the MIPS and {\em SPIRE} photometry of the
galaxy, Engelbracht et al. (2010) find that the central region of NGC
1097 is 22\% warmer than the disk.

The mid- and far-IR images of the ring in Fig~\ref{fig:ring} show
similar structures.  The ring is continuous (i.e. no obvious gaps)
with a series of bright knots.  The same knots are visible in each
PACS image, although at 160 \micron\ they are not well-resolved. At 70
and 100 \micron, the surface brightness of the ring varies by less
than $\pm 15$\% about the mean on 600 pc scales. The variations at 24
\micron\ on the same spatial scales are $\pm 25$\%.  The similarities
from mid- to far-IR suggest that dust temperatures are not varying
substantially in the ring, which we quantify below.  

It is interesting to note that the same pattern of bright knots is not
observed in carbon monoxide (CO) (shown in panel h of
Fig~\ref{fig:ring}) or other dense gas tracers at comparable
resolution \citep{kohno03,hsieh08}.  Instead the CO intensity peaks
near the contact points and is much fainter over the rest of the ring.
The differences between the far-IR and CO emission may be due to
different CO excitation mechanisms in the shocked gas near the contact
points or by the consumption and/or dissociation of molecular gas by
star-formation events shortly downstream from the contact points.
Three of the bright knots are also prominent in 3.5 cm radio continuum
(as shown in panel f of Fig~\ref{fig:ring}).  \citet{beck05}
showed that the radio knots have a flatter radio spectral index than
the rest of the ring, most likely due to either a contribution from
free-free emission from H II regions or synchrotron emission from
young supernova remnants (SNRs), which has an intrinsically
flatter spectrum. Because young SNRs will heat only a small fraction of
the dust, an enhancement of thermal radio continuum and dust heating
in and around H II regions may be the best explanation for the origin
of the coincident bright radio and far-IR knots.   

\begin{figure}
\begin{center}
\scalebox{0.4}{{\includegraphics{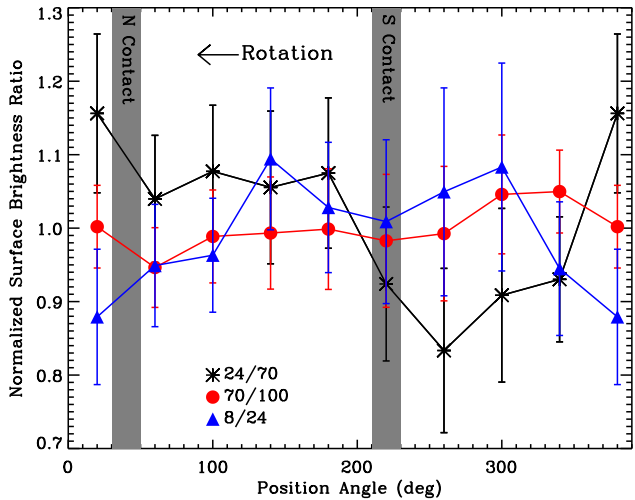}}}
\caption{Azimuthal variations of the mid- and far-IR band ratios
with azimuth in the ring. The x-axis shows the position angle from North
through East (increasing counter-clockwise from North on
Fig~\ref{fig:ring}). Note that the first point is repeated at
380$^{\circ}$.  The band ratios are normalized to their mean
values: 0.487, 0.069, and 0.776 for the 8.0/24, 24/70 and 70/100
ratios, respectively. The contact points of the dust lanes and the
direction of rotation for the ring are labeled on the
plot.\label{fig:randaz}}
\end{center}
\end{figure}

In Fig~\ref{fig:randaz} we show the mid- and far-IR band ratios as
a function of azimuthal angle. The surface brightness was measured in
9 azimuthal bins with inner and outer radii of 5 and 15\arcsec\ to
adequately sample the PSF out to 100 \micron.  The largest variations
are in the 24/70 ratio, which peaks shortly downstream from the
northernmost contact point, and varies by $\pm 15$\%.  The 70/100
ratio varies by less than $\pm 5$\% around the ring.  If there is a
well-defined age gradient along the ring as predicted by the ``pearls
on a string'' model, one might expect a gradient in dust temperatures
moving away from the contact points.  In the youngest star-forming
regions, the radiation field will be more intense due to the presence
of the most massive and short-lived stars and the regions will be
more compact.  Both of these effects lead to hotter dust temperatures.
For instance, \citet{groves08} modeled the spectra of the H II regions
plus surrounding photo-dissociation region for star clusters with ages
between 0.1$-$10 Myr.  They find that for a cluster mass of $\sim
10^5$ \msun\ and an ISM pressure of P/k $\sim 10^6$ K cm$^{-3}$
\citep[approximately what has been deduced for the ISM in the
circumnuclear ring by][]{hsieh08}, the 24/70 and 70/100 band ratios 
decrease by 90 and 70\% (factors of $\sim 7$ and 3, respectively) as
the cluster ages from 1 to 10 Myr.  The \citet{groves08} models
represent an upper bound to the band ratio variation we could expect
if the ring was comprised solely of a well-defined sequence of aging
clusters between 1$-$10 Myr old.  

That we do not see large mid- and far-IR band ratio 
gradients does not rule out the existence of ``pearls on a string''
in favor of ``popcorn'' in NGC 1097, however.  It may be the
case that dust in the ring is predominantly heated by the radiation
field from older stars (e.g. B stars with lifetimes of 10-100 Myr),
which are uniformly distributed around the ring after a number of
revolutions.  Stellar population studies of the central kpc of
NGC 1097 have shown that intermediate age stars make a considerable 
contribution to the UV radiation field in the vicinity of the ring
\citep[e.g.][]{bonatto98}. In this situation, the variation due to an
age gradient would be diluted depending on the relative contribution
of young clusters to the total dust heating.  In addition, one might
expect that given the fast dynamical time in the ring that local
enhancements of dust heating are quickly wiped out.  Stars and
interstellar matter in NGC 1097's ring traverses the distance
between the two contact points in $\sim 9$ Myr.  Even assuming that
the cluster formation happens instantaneously after entering the ring,
there will still be an abundance of massive stars by the time the
cluster crosses half the ring.   

\section{Limits on the nuclear flux in the far-IR}

\begin{figure}
\begin{center}
\scalebox{0.4}{{\includegraphics{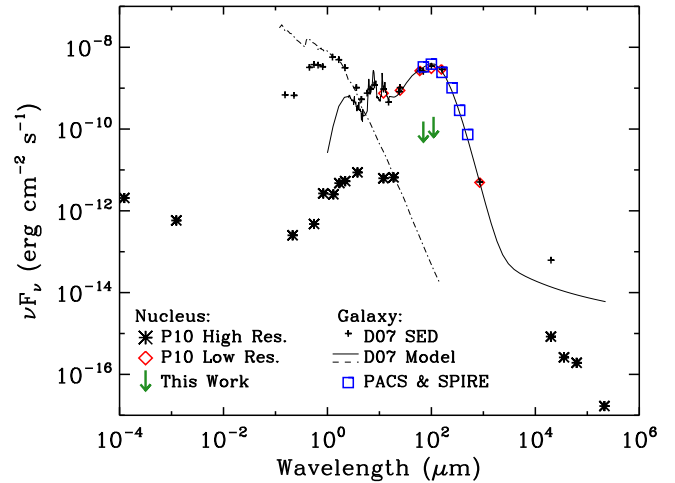}}}
\caption{SED of the nuclear region from the compilation of
\citet{prieto10} (P10) and the SED of the whole galaxy from
\citet{dale07} (D07).  The black crosses show the D07 SED overlaid
with their best-fit model to the MIPS observations.  The blue squares
show PACS and {\em SPIRE} fluxes for NGC 1097 from KINGFISH observations
(see Engelbracht et al. 2010 for discussion of the {\em SPIRE}
observations). The black asterisks show observations at high spatial
resolution (on the order of 1\arcsec\ or less).  The red points are
lower resolution IR measurements of the central region which are
dominated by the flux from the starburst ring.  The green arrows show
the new limits we can place on the nuclear flux using our PACS
observations.\label{fig:sed}}
\end{center}
\end{figure}

High resolution imaging at mid- and near-IR wavelengths of the nucleus
of NGC 1097 shows an unresolved central point source
\citep{prieto05,mason07} which contains a nuclear starburst
\citep{storchi-bergmann05} and the AGN.  No observations
can yet resolve the AGN or central starburst, but it is still possible to
distinguish the contributions of the different sources to some degree.
\citet{mason07}, for instance, found that the 12 and 18 \micron\
emission arises primarily from dust heated by the nuclear starburst
rather than the AGN torus.  In NGC 1097, previous far-IR flux limits
for the nucleus were dominated by the starburst ring and provide
limited information about the nuclear starburst or the AGN.  With our
{\em Herschel} observations, we can place limits on the flux arising
in the central $\sim 600$ pc.  We use the PACS PSF observations of
Vesta with a 20\arcsec s$^{-1}$ scan speed scaled to match the peak
intensity of the point source we see in the center.  Our scaled-PSF
photometry is possible at 70 and 100 \micron, but not at 160 \micron\
where the central source is not well-resolved.  The best scaled PSF
has a total flux of 3.5 and 7.3 Jy at 70 and 100 \micron.  Without
more detailed modeling of the nuclear region these measurements should
only be considered upper limits.  However, they improve constraints on
the nuclear flux by more than an order of magnitude as shown in
Fig~\ref{fig:sed}, which presents the SED of the nucleus  from a
compilation by \citet{prieto10}.  

\section{Conclusions}

We have presented {\em Herschel} PACS observations from KINGFISH of
the inner kpc of the barred spiral galaxy NGC 1097.  These are the
first observations to resolve a starburst ring at wavelengths probing
the peak of the dust SED.  We show a comparison of the ring in a
variety of tracers and find similar bright knots in the mid- and
far-IR and radio continuum.  These knots do not correspond to the same
knots traced by CO.  We find modest variation azimuthally in the
mid- and far-IR band ratios suggesting that either there is no
azimuthal age gradient, as would be predicted by the ``pearls on a
string'' mode of star-formation, that dust heating is dominated by an
older stellar population and/or that the dust heating variations get
quickly erased over the short ring orbital period ($\sim 18$ Myr).
Finally, we place an order-of-magnitude tighter constraint on the
far-IR emission originating in the central $\sim 600$ pc of the
galaxy.

\acknowledgements

The authors thank R. Beck for the radio continuum data and P.-Y.
Hsieh for the CO data.  K.S. would like to thank G. van de Ven and L.
Burtscher for useful discussions. PACS has been developed by a
consortium of institutes led by MPE (Germany) and including UVIE
(Austria); KU Leuven, CSL, IMEC (Belgium); CEA, LAM (France); MPIA
(Germany); INAF- IFSI/OAA/OAP/OAT, LENS, SISSA (Italy); IAC (Spain).
This development has been supported by the funding agencies BMVIT
(Austria), ESA-PRODEX (Belgium), CEA/CNES (France), DLR (Germany),
ASI/INAF (Italy), and CICYT/MCYT (Spain).

%\bibliographystyle{aa}
%\bibliography{/Users/karin/Desktop/sdp_paper/for_submission/sdp}

\end{document}